\documentclass[12pt]{iopart}
\usepackage{epsfig}
\begin{document}

\title[Strangeness, Quasi-Quarks \& Lattice QCD]{Lattice QCD Results on 
Strangeness and Quasi-Quarks in Heavy-Ion Collisions
\vskip-2.9cm\hfill\small hep-ph/0605254, TIFR/TH/06-11\vskip 2.6cm
}

\author{Rajiv V. Gavai\footnote[3]{Speaker at the conference}
and Sourendu Gupta}

\address{Department of Theoretical Physics, 
Tata Institute of Fundamental Research, \\
Homi Bhabha Road, Mumbai 400 005, India}

\begin{abstract}

Fluctuations of conserved quantities in heavy-ion collisions have been argued
to be diagnostic tools for the nature of the produced phase.  These can be
related to the predictions of quark number susceptibilities (QNS) from lattice
QCD. Using the diagonal QNS, we extracted the Wr\'oblewski parameter in a
dynamical QCD computation. Our results on the cross correlations $\chi_{BQ}$,
$\chi_{BY}$, $\chi_{BS}$ and $\chi_{QY}$ allow us to  explore the charge and
baryon number of objects that carry flavour.  We present evidence that in the
high temperature phase of QCD the different flavour quantum numbers are excited
in linkages which are exactly the same as one expects from quarks. 

\end{abstract}



\maketitle

\section{Introduction}
As exciting experimental results keep pouring in from the Relativistic Heavy
Ion Collider (RHIC) in BNL, New York, USA, the task of detecting the
quark-gluon plasma (QGP) and establishing its various properties becomes more
and more pressing.  One needs therefore to look at as many signatures in as
different a variety of aspects as possible.  This has, of course, been going on
both theoretically and experimentally. Fluctuations \cite{AJ} in conserved
quantities such as the baryon number $B$, electric charge $Q$ or strangeness
$S$ have been proposed as promising signals of QGP.  Similarly, since very
early days of relativistic heavy ion collisions, enhancement of strangeness
production  \cite{RM} has also been regarded as a useful indicator.    

The original proposals were usually based on simple model considerations as was
the case for  many other signals.  The key fact employed there was that the
up/down and strange quark masses are smaller than the expected transition
temperature whereas the masses of the low lying hadrons, except for pion, are
significantly larger.  While these arguments on the differences between the two
phases are qualitatively appealing, one has to face quantitative questions of
details for any meaningful comparison with the data.  Since the temperature of
the plasma produced in RHIC, or even LHC, may not be sufficiently high for
simple ideal gas pictures to be applicable, numerical estimates or indeed even
the degree of utility of the proposal, could be misleading.  Information
obtained directly from the underlying theory, quantum chromodynamics (QCD)
appears desirable, and the lattice formulation of QCD can potentially deliver
that.

A variety of aspects of the strangeness enhancement have been studied and 
many different variations have been proposed.   One very useful way of
looking for strangeness enhancement is the Wr\'oblewski parameter \cite{Wr}.
Defined as the ratio of newly created strange quarks to light quarks, 
\begin{equation}
\lambda_s = \frac{2\langle s\bar s\rangle}{\langle u\bar u+d\bar d\rangle}
\end{equation}
the Wr\'oblewski parameter has been estimated for many processes using a hadron
gas fireball model \cite{BH}.  An interesting finding from these analyses is
that $\lambda_s$ is around 0.2 in most processes, including proton-proton
scattering, but is about a factor of two higher in heavy ion collisions. An
obvious question one can ask is whether this rise by a factor of two can be
attributed to the strangeness enhancement due to quark gluon plasma and if yes,
whether this can be quantitatively demonstrated.  We show below how quark
number susceptibilities, obtained from simulations of lattice QCD, may be
useful in answering these questions.

An interesting property that the Wr\'oblewski parameter possesses is that it is
a robust quantity.  Indeed, we \cite{us} have recently argued that the ratios
of susceptibilities, $C_{K/L}$, defined by 
\begin{equation}  
C_{K/L} \equiv \frac{\chi_K}{\chi_L} = \frac{\sigma^2_K}{\sigma^2_L}, 
\end{equation}  
where $\chi_K$ and $\chi_L$ are quark number susceptibilities (QNS) for the
conserved quantum numbers $K$ and $L$ and $\sigma_K$ and $\sigma_L$ are the
corresponding experimentally measured variances, are robust variables in the
high T Phase.  This is so both theoretically and experimentally, provided the
two variances are obtained under identical experimental conditions and after
removing counting (Poisson) fluctuations.  We will show below our lattice QCD
results in partial support of this claim.

We employed such robust variables to address the important outstanding question
of the nature of plasma (QGP) excitations.  From several lattice QCD
investigations, it has been known for a long time that the straightforward
perturbation theory fails to describe the lattice data for equation of state
for $T \ge T_c$.  Various resummation schemes, and phenomenological models have
been tried to understand the nature of the QGP in this region.  For $T/T_c \ge
3-5$, lattice results on entropy density seem to be in agreement with such
modified weak coupling pictures. Quark number susceptibilities provide an
independent check on them, and yielded a strong support for these ideas.
Screening mass determinations also suggested an agreement with an ideal Fermi
gas of quarks for  $T \ge 2 T_c$.  Here we advocate a more direct approach.
Creating an excitation of quantum number $K$, one can ask what else, e.g., like
another quantum number $L$, does it carry.  We address this by using the ratios
of off-diagonal susceptibility $\chi_{KL}$ with $\chi_L$  :
\begin{equation}  
C_{(KL)/L} = \frac{\langle KL\rangle
-\langle K\rangle\langle L\rangle}{ \langle L^2\rangle-\langle L\rangle^2}~.
\end{equation}

\section{Quark Number Susceptibilities}

Let us first describe in this section the way one obtains various QNS
from first principles using the lattice formulation of QCD.
Assuming three flavours of quarks, and denoting by $\mu_f$ the
corresponding chemical potentials, the QCD partition function is
\begin{equation}
{\cal Z} =  \int D U~~ \exp(-S_G)  \prod_{f=u,d,s} 
{\rm Det}~M(m_f, \mu_f)~~.
\label{zqcd}
\end{equation}
Note that the quark mass and the corresponding chemical potential enter only
through the Dirac matrix $M$ for each flavour.  We employ the standard Wilson
form for the gluonic action $S_G$ and staggered fermions (with the usual
square root method) to describe the quark matrix $M$.  The formalism is,
however, general and any change in either affects only the detailed
expressions.  The final physical results are expected to remain the same for
sufficiently small lattice spacing $a$.

Defining $\mu_0 = \mu_u + \mu_d + \mu_s$ and $\mu_3 = \mu_u - \mu_d$, the
baryon and isospin densities and the corresponding susceptibilities can be
obtained as:  
\begin{equation}
\qquad \qquad n_i =  \frac{T}{V} {{\partial \ln {\cal Z}}\over{\partial \mu_i}}, \qquad
\chi_{ij} =  \frac{T}{V} {{\partial^2 \ln {\cal Z}}\over{\partial \mu_i 
\partial \mu_j} }  ~.
\label{nchi}
\end{equation}
Similarly, Charge ($Q$), Hypercharge ($Y$), Strangeness ($S$) susceptibilities
can be defined.  QNS in \eref{nchi} are crucial for many quark-gluon plasma
signatures which are based on fluctuations in globally conserved quantities
such as baryon number or electric charge.  Theoretically, they serve as an
important independent check on the methods and/or models which aim to explain
the large deviations of the lattice results for pressure $P$($\mu$=0) from the
corresponding perturbative expansion.  Here we will be concerned with i) 
extending our earlier quenched results on the Wr\'oblewski parameter 
to full QCD with two dynamical quarks using our proposal \cite{us1} to
estimate it from the QNS:
\begin{equation}
\lambda_s = {2 \chi_s \over { \chi_u +\chi_d}}~, 
\label{wrob}
\end{equation}
and ii) demonstrating the relevant degrees of freedom in QGP above $T_c$.

In order to use \eref{wrob} to obtain an estimate for comparison with
experiments or for ii) above, one needs to compute the corresponding quark
number susceptibilities on the lattice first and then take the continuum limit.
All susceptibilities can be written as traces of products of $M^{-1}$ and
various derivatives of $M$ with respect to $\mu$.  With $m_u = m_d$, diagonal
$\chi_{ii}$'s can be written as

\begin{eqnarray}
\label{chiexp1}
\chi_{00} &=& \frac{T}{2V} [ \langle {\cal O}_2(m_u) + \frac{1}{2} {\cal O}_{11}(m_u) \rangle ] \\ 
\label{chiexp2}
\chi_{33} &=& \frac{T}{2V} ~~ \langle {\cal O}_2(m_u) \rangle  \\ 
\label{chiexp3}
\chi_{ss} &=& \frac{T}{4V} [ \langle {\cal O}_2(m_s) + \frac{1}{4} {\cal O}_{11}(m_s) \rangle ] ~~.
\end{eqnarray} 

\noindent
Here ${\cal O}_2 = {\rm Tr}~M^{-1}_u M_u'' - {\rm Tr} ~M^{-1}_u M_u'M^{-1}_u
M_u'$, and $ {\cal O}_{11}(m_u) = ({\rm Tr}~M^{-1}_u M_u' )^2$, with the
prime(s) denoting first(second) derivative of $M(\mu_i)$ with respect to
$\mu_i$ at $\mu_i = 0$.  The traces are estimated by a stochastic method: $
{\rm Tr}~A = \sum^{N_v}_{i=1} R_i^\dag A R_i / 2N_v$, and $ ({\rm Tr}~A)^2 = 2
\sum^{L}_{i>j=1} ({\rm Tr}~A)_i ({\rm Tr}~A)_j/ L(L-1)$, where $R_i$ is a
complex vector from a set of $N_v$, subdivided further in L independent sets.

Higher order susceptibilities are defined by 
\begin{equation}
   \chi_{fg\cdots} =
      \frac TV\frac{\partial^n \log Z}{\partial\mu_f\partial\mu_g\cdots} =
      \frac{\partial^n P}{\partial\mu_f\partial\mu_g\cdots}~.
\end{equation}
These are Taylor coefficients of the pressure $P$ in its expansion in $\mu$.
All of these can be written as traces of products of $M^{-1}$ and various 
higher derivatives of $M$.  These too can be evaluated using the Gaussian noise
technique described above, but with increasingly larger number of
vectors for progressively higher orders.  One can build the expansion
order by order to obtain information on the critical point of QCD
in the $T-\mu$ plane, as done, for example in \cite{usc}.

\begin{figure}[htb]
\begin{minipage}{0.49\textwidth}
\epsfxsize=7.2cm
\epsfbox{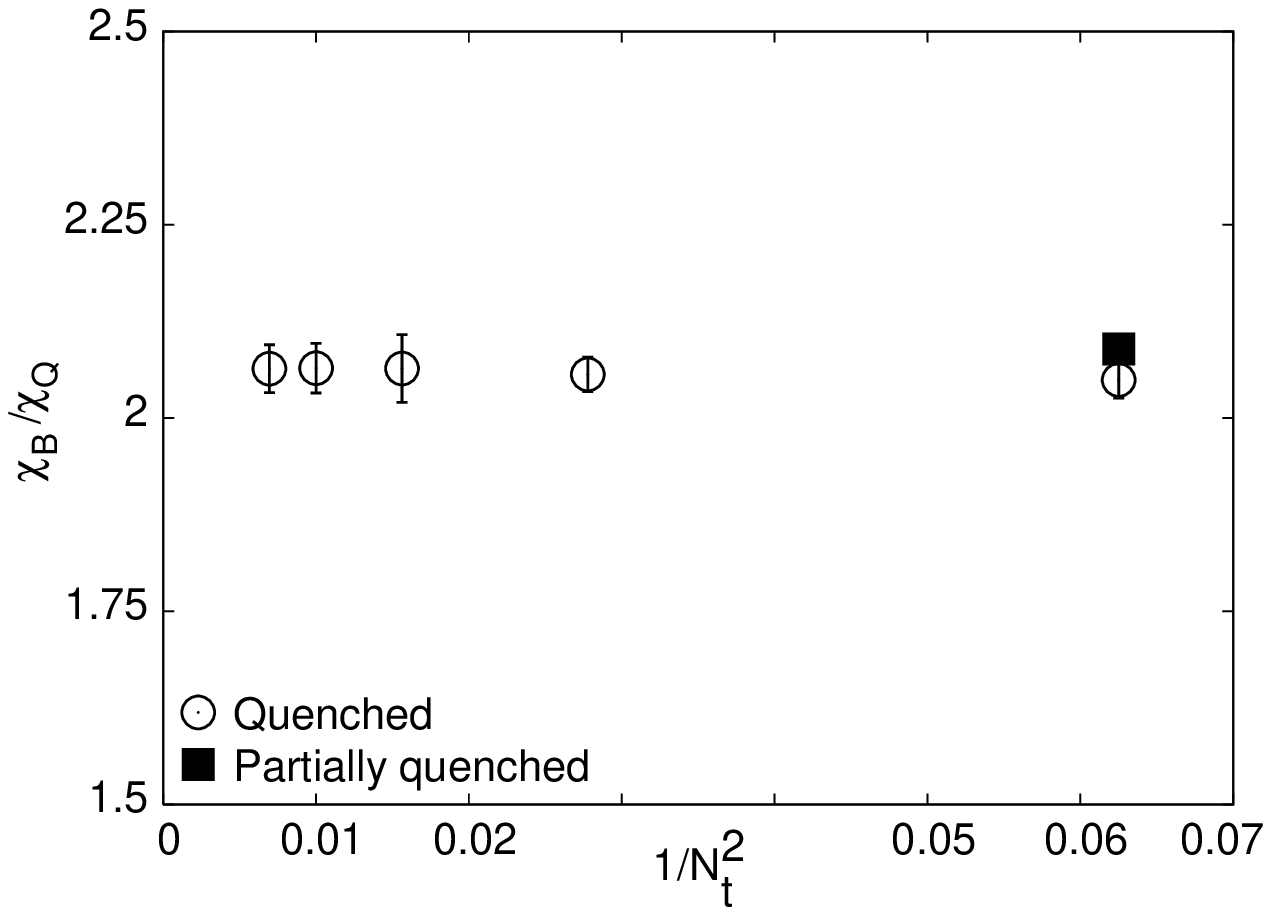}
\end{minipage}
\begin{minipage}{0.49\textwidth}\raggedright
\epsfxsize=7.2cm
\epsfbox{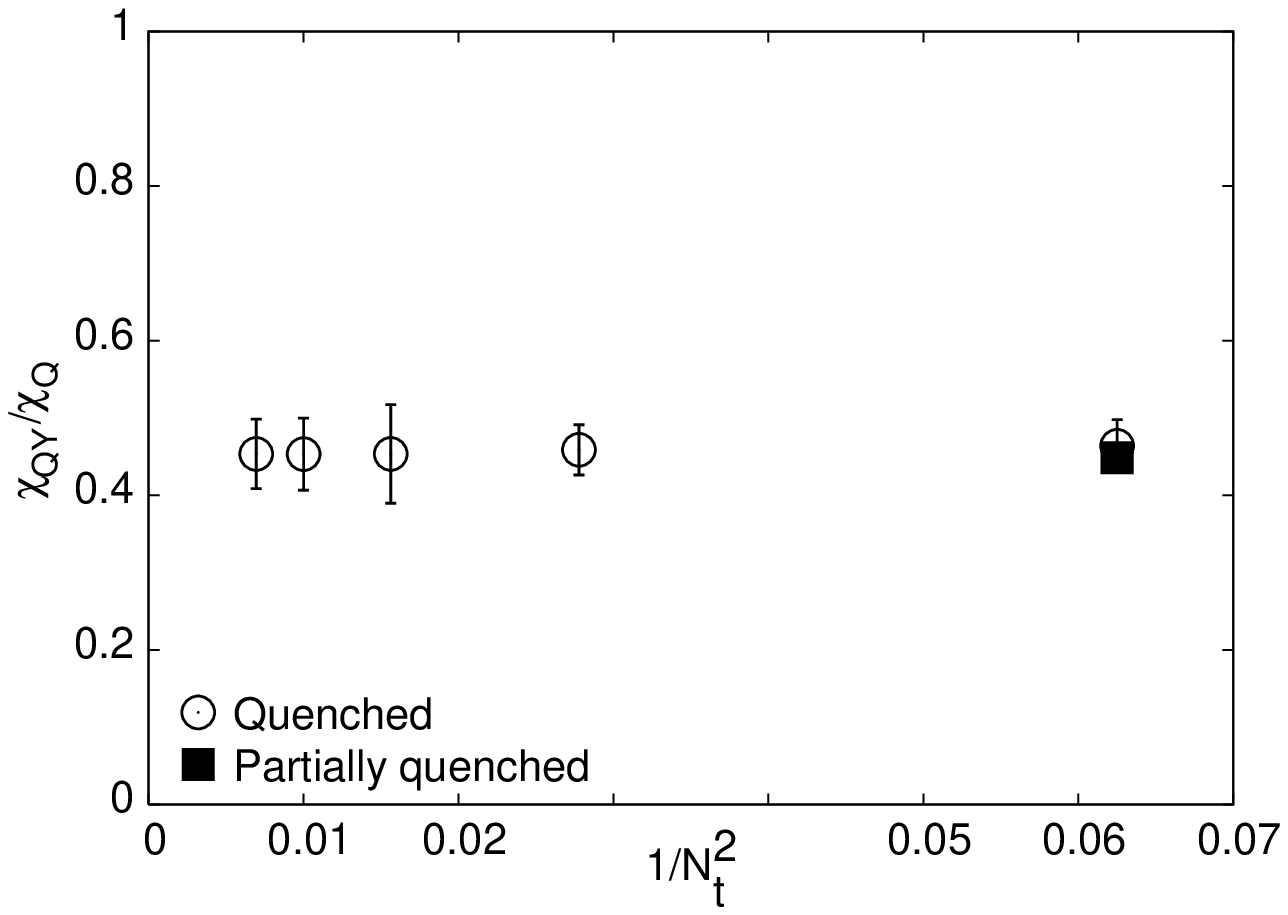}
\end{minipage}
\label{fig1}
\caption{ $C_{B/Q}$ (left) and $C_{(QY)/Q}$ (right) exhibited as a function 
of $1/N_t^{-2} \propto a^2$ at $T=2T_c$, where $a$ is the lattice spacing.
Ratios of QNS are seen to be robust observables--- being insensitive 
to both changes in lattice spacing $a$ and the sea quark content of QCD.  }
\end{figure}


Figure 1 displays our results \cite{us} for the  ratios $C_{B/Q}$ (left panel)
and $C_{(QY)/Q}$ (right panel) as a function of (square of) the lattice spacing
$a$ at $2T_c$, where $T_c$ is the transition temperature. The open symbols are
for quenched QCD while the filled ones are for simulations with two flavours of
light dynamical quarks.  While the quenched results were obtained from a
reanalysis of our earlier simulations \cite{us1}, the data for the $N_f=2$ full
QCD, also denoted as partially quenched to distinguish from the real world case
of two light and one heavy strange quark,  were obtained \cite{usc} on
$4\times16^3$ lattices.  The configurations were generated with a bare sea
quark mass $m=0.1T_c$, which gives $m_\pi=0.3 m_\rho$.  The setting of scale,
the parameters employed and the statistics are detailed in Ref. \cite{usc}. To
that set of data with $T/T_c$ = $0.75\pm0.02$, $0.8\pm0.02$, $0.85\pm0.01$,
$0.9\pm0.01$, $0.95\pm0.01$, $1.00\pm0.01$, $1.045\pm0.01$, $1.25\pm0.02$,
$1.65\pm0.06$ and $2.15\pm0.10$, we added two more sets--- 55 configurations
separated by more than two autocorrelation times at $T/T_c=0.975\pm0.010$
(i.e., $\beta=5.2825$) and 86 configurations, similarly spaced, at
$T/T_c=1.15\pm0.01$ (i.e., $\beta=5.325$).

In both the quenched and the partially quenched case and for both figures, the
light valence quark mass $m_{\rm v}$, appearing in
\eref{chiexp1}-\eref{chiexp3}, is $0.03T_c$ and the strange valence quark mass
is $T_c$.  Both these ratios, and similar other ratios, display the property
that $r(a)=r+{\cal O}(a^n)$ with $n>2$, implying very good scaling behaviour.
Furthermore, the results for the quenched and the partially quenched case on
the $N_t=4$ lattice differ by a few per cent only, although $T_c$ in the two
cases differ by a factor of 1.6.   These two aspects of these ratios, very
little dependence on the sea quark mass and the lattice spacing $a$, prompted
us to term them as robust observables.  Such robust ratios $r$ can be reliably
extracted on the smallest lattice already, which is very useful for the
computationally expensive simulations with the dynamical quarks.  One expects
this robustness to persist at other temperatures above $T_c$ as well.  At lower
temperatures, all susceptibilities, including the off-diagonal ones, are of
similar order and such robustness may not persist. It would, nevertheless, be
interesting to test it carefully.

\begin{figure}[htbp]\begin{center}
\epsfig{height=7cm,file=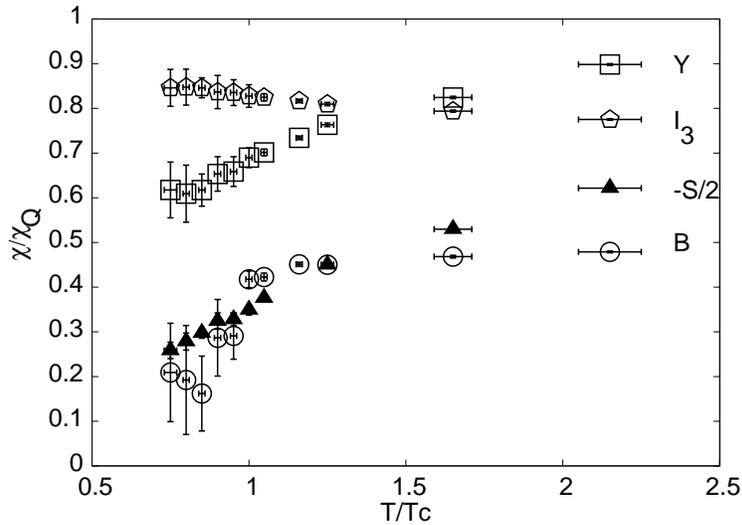}
\caption{Some robust predictions of fluctuation measures from QCD. }
\label{fig2}\end{center}
\end{figure}
\section{Robust Predictions :  Wr\'oblewski Parameter}

Figure 2 exhibits some robust predictions of fluctuation measures from QCD,
obtained from our dynamical simulations with light $u$ and $d$ quarks.  Shown
are the ratios $C_{X/Q}$ for the quantum number $X$ indicated in the figure.
The $X$ = $-S$ case is normalized by an extra factor of 2 to show it on the
same plot.  Note that the hierarchy seen in Figure 2 is itself a robust
feature.  Our results indicate that experimental studies of $C_{S/Q}$ and
$C_{B/Q}$ are most promising in terms of distinguishing between the two
phases of QCD, because they exhibit the largest changes in going from the
hadronic phase to QGP.

Recall that the strange quark susceptibility was obtained from the same
simulations by simply choosing $m_{\rm v}/T_c = 1$. Using \eref{wrob},
$\lambda_s(T)$ can then be easily obtained. Earlier we \cite{us1} had
extrapolated $\lambda_s(T)$ in quenched QCD to $T_c$ by employing simple
ans\"atze.  This became necessary in order to avoid any effects of the order
of the phase transition; quenched QCD has first order transition while the
$N_f=2$ has a lot smoother transition.  While the resultant $\lambda_s(T_c)$
in quenched QCD displayed an impressive agreement \cite{us1} with the results
obtained from the analysis of the RHIC and SPS data in the fireball
model\cite{BH}, clearly a cleaner determination with much better control of
systematic errors needs its determination at $T_c$ in full QCD.  Noting that
$\lambda_s(T)$ is a robust observable as well, one can do so from our $N_t=4$
simulations, at least for $T \ge T_c$.

Left panel of Figure 3 shows the Wr\'oblewski parameter for full QCD as a
function of $T/T_c$. Its value at $T_c$ is seen to be $\lambda_s\approx0.4$, in
agreement with the value of extracted from experiments, when the freeze-out
temperature is close to $T_c$ \cite{BH}.  It is also a pleasant fact that at
lower temperatures the ratio keeps decreasing.  It is rather sensitive to the
quark masses at those temperatures.  The data sets  A, D and C in the right
panel of Figure 3 display $\chi_{BY}/\Delta_{us}^2$, i.e, the susceptibility
$\chi_{BY}$ divided by the square of the mass difference $\Delta_{us} = m_s
-m_u$, for  $m_s/T_c$ = 0.1, 0.75 and 1 and $m_u/T_c =0.03$.  It is easy to
show \cite{us} that such cross susceptibilities are zero in the limit of equal
quark masses, and are $\propto \Delta_{us}^2$. Strong kinematic effects are
clearly visible near $T_c$.  It remains to be seen whether a realistic set of
quark of quark masses, corresponding to the physical pion and Kaon masses,
brings $\lambda_s$ rapidly down by a factor of two or so below $T_c$.  

\begin{figure}[htb]
\begin{minipage}{0.49\textwidth}
\epsfxsize=7.2cm
\epsfbox{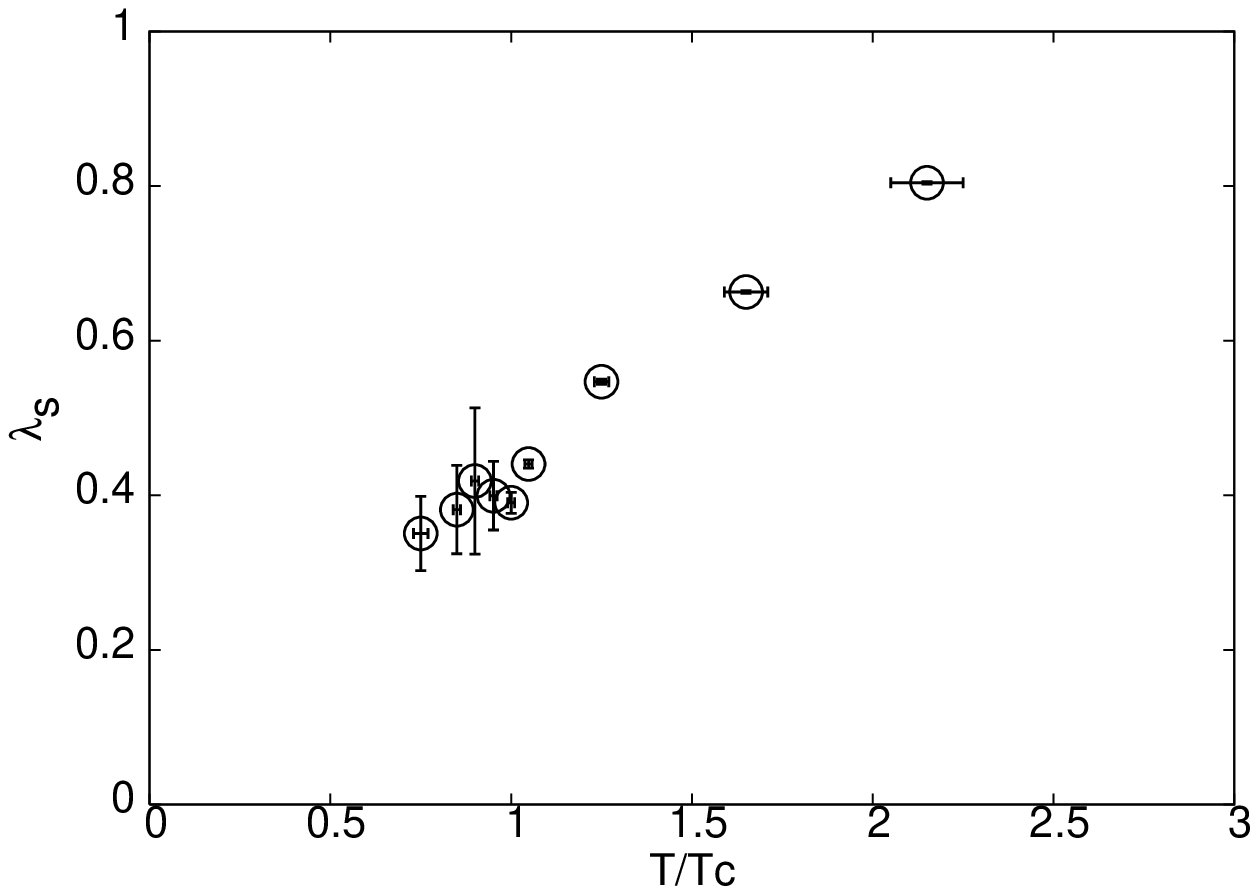}
\end{minipage}
\begin{minipage}{0.49\textwidth}\raggedright
\epsfxsize=7.2cm
\epsfbox{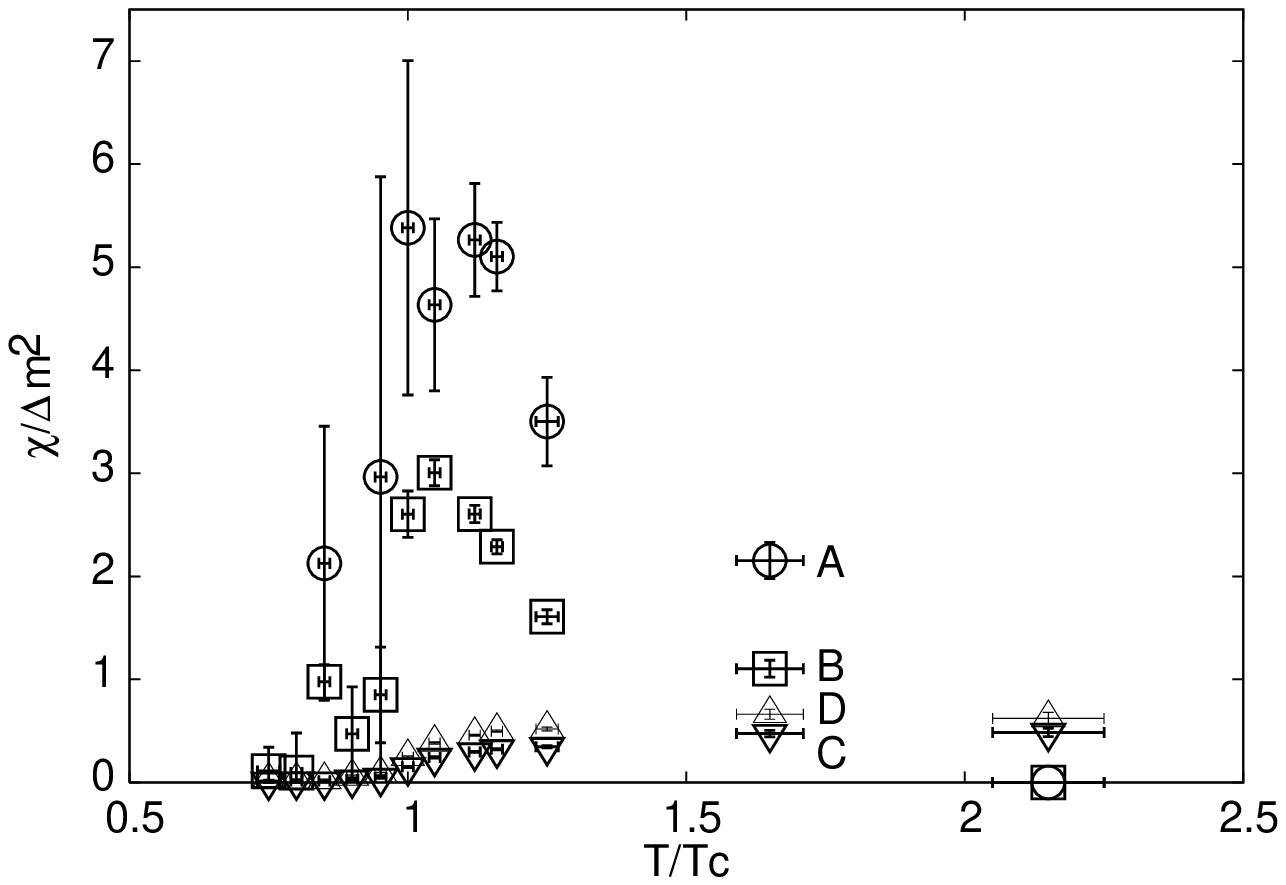}
\end{minipage}
\label{fig3}
\caption{$\lambda_s$ as a function of temperature for QCD with two light
dynamical quarks (Left panel) and various flavour symmetry breaking matrix 
elements (defined in the text) as a function of $T/T_c$ (Right panel).}
\end{figure}


The above mentioned nice agreement needs to be treated cautiously, however, in
view of the various approximations made in equating the robust ratio determined
on the lattice with the Wr\'oblewski parameter extracted from the data.  Let us
list them in order of severity.

\begin{itemize}

\item The experiments at RHIC and SPS have nonzero albeit small $\mu$ whereas
the above lattice result used $\mu=0$. Based on both lattice QCD and fireball
model considerations, $\lambda_s$ is expected to change very slowly for small
$\mu$.  This can, and should, be checked by direct simulations.

\item Lattice simulations yield real quark number susceptibility whereas for
particle production its imaginary counterpart is needed.  Assuming that the
characteristic time scale of plasma are far from the energy scales of strange
or light quark production, one can relate \cite{book} the two to justify
the use of lattice results in obtaining $\lambda_s$.   Observation of spikes 
in photon production may falsify this assumption. 

\end{itemize}

\section{Flavour Carriers : Quasi-quarks ?}

Unlike the gluons, quarks carry flavour such as electric charge or strangeness.
Flavour in quark sector can potentially assist in identification of relevant
degrees of freedom in QGP just above $T_c$. One can look for correlations, by
exciting one quantum number and looking for the presence of another.  Choosing,
baryon number and strangeness as these quantum numbers, Koch, Majumder and
Randrup \cite{koch} introduced the variable
\begin{equation}
   C_{BS} = -3C_{(BS)/S} = 1 + \frac{\chi_{us}+\chi_{ds}}{\chi_s}~,~
\label{corr}
\end{equation}
in order to distinguish between bound state QCD --sQGP \cite{bqcd}-- and the
usual picture of the excitations in the plasma phase of QCD.  This is expected
to have a value of unity if strangeness is carried by (ideal) quarks, since
$S=1$ always comes linked with $B=-1/3$.  In \cite{koch} it was shown that
bound state QGP gives a value of $C_{BS}\approx2/3$ (for $T>T_c$).
Charge and strangeness correlation offers another similar 
possibility of being unity, if strangeness is carried by quarks :
\begin{equation}
 C_{QS} = 3C_{(QS)/S} = 1 - \frac{2\chi_{us}-\chi_{ds}}{\chi_s}~. 
\label{corrqs}
\end{equation}

\begin{figure}[htbp]\begin{center}
\epsfig{height=7cm,file=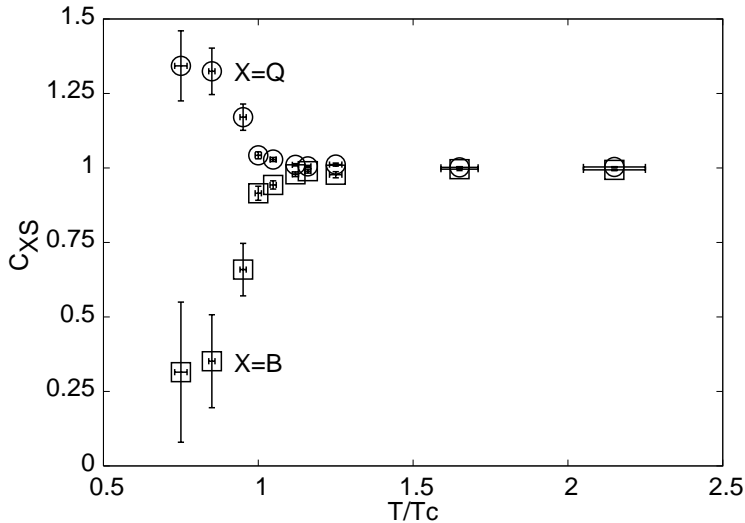}
\caption{$C_{BS}$ and $C_{QS}$, as functions of $T/T_c$.
   The quark masses used are $m_{ud}=0.1T_c$ and $m_s=T_c$, although in the
   high temperature phase there is no statistically significant dependence
   on the quark masses.}
\label{fig4}\end{center}
\end{figure}

In Figure 4,  we show the first lattice QCD results \cite{us} on $C_{BS}$ and
$C_{QS}$ from our dynamical two light quark simulations.  Note that while both
are different from unity below $T_c$, they become close to unity immediately
above $T_c$, strongly indicating that unit strangeness is carried in QGP by
objects with baryon number -1/3 and charge 1/3 very close to $T_c$ and beyond.
The trends for $T < T_c$ as well as the order of magnitude estimated  
\cite{koch} in the hadron gas models are in agreement with our lattice results.
We varied the strange quark mass, $m_s/T_c$,  between 0.1 and 1.0 and found
that for $T \ge T_c$ it does not alter the value, $\approx 1$, or the visible
$T$-independence.  A natural explanation of the $T$-behaviour is provided
if strange excitations with nonzero baryon number become lighter at $T_c$. 
Furthermore, the $T$-independence suggests dominance of a single such
excitation.  Similar results were also obtained in \cite{us} in the light quark 
sector, from e.g., $C_{(BU)/U}$ and $C_{(QU)/U}$,  linking $u$-flavour 
carriers to $B=1/3$ and $Q=2/3$ objects.

A possible interpretation of the observed very small deviations from unity of
$C_{BS}$ and $C_{QS}$ is as follows.  Colour interactions merely dress up
quarks. These effects can be computed in weak coupling limit, i.e, at high
$T$, where the flavour linkages of quarks are natural.  Close to $T_c$, the
flavour linkages remarkably continue to persist as before, but they are no
longer computable as the coupling is presumably not weak.   This leads us to
term them therefore as quasi-quarks.

\section{Summary}

Quark number susceptibilities which can be obtained from first principles using
lattice QCD contribute substantially to the physics of RHIC signals.  Since the
ratios of quark number susceptibilities, $C_{A/B}$ were shown to depend rather
weakly on the lattice spacing and the sea quark content of QCD in the high
temperature phase, they are theoretically robust variables.  Consequently, 
one can compute them reliably relatively easily on small lattices, as we did.
Being ratios, experimentally they are likely to be free of various systematic
errors and thus again robust experimental observables.  We presented first full
(two light dynamical quarks) QCD results for the Wr\'oblewski Parameter
$\lambda_s (T)$.  Near $T_c$, these are found to be in agreement with the RHIC
and SPS results.  Being a robust observable, small lattice cut-off effects
expected in $\lambda_s$.

We demonstrated that the high temperature phase of QCD essentially consists of
quasi-quarks by exploiting the flavour linkages of quarks. In particular, we
showed that unit strangeness is carried by an object with baryon number
$-1/3$ and charge $1/3$, as seen in Figure 4. Moreover, this correlation does
not depend on the strange quark mass even when it is as large as $T_c$.
Similarly, in the light quark sector one finds that u and d quantum numbers are
not produced together, and that the u flavour is carried by excitations with
baryon number $+1/3$ and charge $+2/3$, whereas the d flavour is carried by
particles with baryon number $+1/3$ and charge $-1/3$.

\section{Acknowledgments}

One of us (RVG) wishes to thank the organizing committee, and Prof. Huan Zhong
Huang in particular, for their kind invitation and the financial support which
enabled him to be a part of the very enjoyable ``Strangeness in Quark Matter
2006'' in UCLA, Los Angeles, USA.  Our computation was carried out on the
Indian Lattice Gauge Theory Initiative's CRAY X1 at the Tata Institute of
Fundamental Research. It is a pleasure to thank Ajay Salve for his
administrative support on the Cray.

\vspace{0.5cm}


\end{document}